\documentclass[preprint2]{aastex}
\usepackage{psfig}
\begin{document}

\title{The Peculiar Type Ia Supernova 1999by: Spectroscopy at Early 
Epochs\footnote{Based on observations obtained at David Dunlap Observatory,
University of Toronto, Canada}}

\author{J. Vink\'o} 
\affil{Department of Optics \& Quantum Electronics, University of Szeged
POB 406, Szeged, H-6701 Hungary}

\author{L.L. Kiss\footnote{Visiting Astronomer, David Dunlap Observatory,
University of Toronto, Canada}, B. Cs\'ak, G. F\H ur\'esz}
\affil{Department of Experimental Physics, University of Szeged,
D\'om t\'er 9, Szeged, H-6720 Hungary}

\author{R. Szab\'o}
\affil{Konkoly Observatory of the Hungarian Academy of Sciences,
Konkoly-Thege \'ut 13-17, Budapest, H-1525 Hungary}

\and

\author{J.R. Thomson, S.W. Mochnacki}
\affil{David Dunlap Observatory, University of Toronto, 
Box 360, Richmond Hill, Ontario, L4C 4Y6 Canada} 


\begin{abstract}
We present medium resolution 
($\lambda / \Delta \lambda \approx 2500$) optical spectroscopy
of SN~1999by in NGC~2841 made around its light maximum. The depth ratio of
the two Si~II features at 5800 \AA\ and 6150 \AA\ being 
$R$(SiII)$\approx 0.63$ at maximum indicates that this SN belongs to the
peculiar, sub-luminous SNe Ia. Radial velocities inferred from
the minimum of the 6150 \AA\ trough reveal a steeper decline of
the velocity curve than expected for ``normal'' SNe Ia, consistent
with the behavior of published $VRI$ light curves.  A revised
absolute magnitude of SN~1999by and distance to its host galaxy
NGC~2841 is estimated based on the Multi-Color Light Curve Shape
(MLCS) method, resulting in $M_V$(max)=$-$18.06 $\pm 0.1$ mag and
$d = 17.1 \pm 1.2$ Mpc, respectively. An approximative 
linear dependence of the luminosity parameter
$\Delta$ on $R$(SiII) is presented.
\end{abstract}


\section{Introduction}

\begin{figure*}
\begin{center}
\leavevmode
\psfig{file=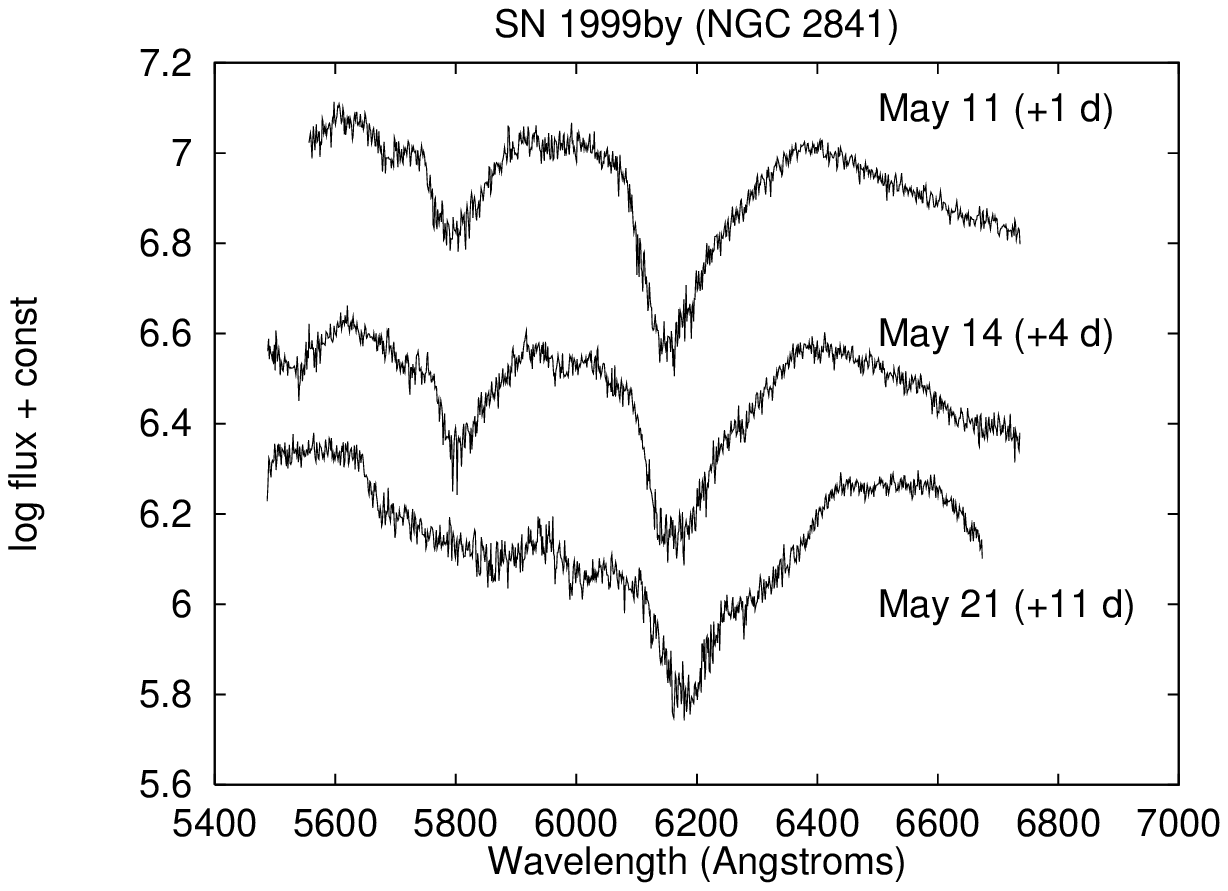,width=10cm}
\psfig{file=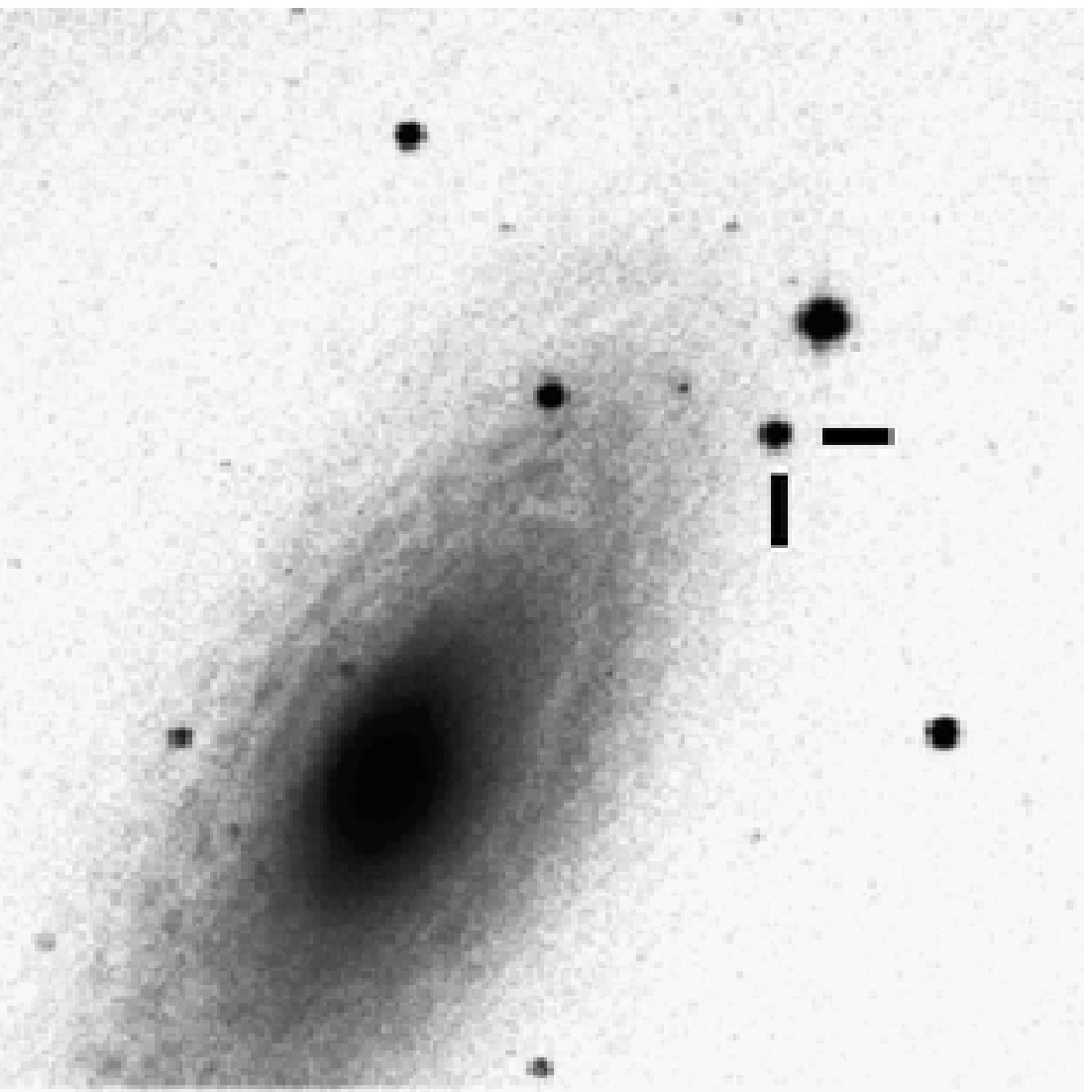,width=6cm}
\caption{Left: spectra of SN 1999by, shifted vertically for better visibility.
Epochs are indicated on the right side of each spectrum.
Right: CCD-image of SN~1999by (marked) taken with the 1m RCC-telescope
at Piszk\'estet\H o station of the Konkoly Observatory on May 21, 1999
(north is up and east is to the left).}
\end{center}
\end{figure*}

In the last decade it became evident that some of the 
observed Type Ia supernovae are intrinsically sub-luminous
compared to the majority of so-called ``normal'' SNe Ia.
SN~1991bg (Filippenko et al., 1992) is now considered 
as their prototype, and the sample of 
well-observed events includes SN~1992K (Hamuy et al., 1994),
SN~1997cn (Turatto et al., 1998) and SN~1998de (Modjaz et al., 2000).
Beside being about 1 magnitude fainter than normal SNe Ia at
maximum, sub-luminous SNe Ia also exhibit peculiar colors and spectra.
For example, the intrinsic (unreddened) $(B-V)_0$ color around
maximum is 0.7 $-$ 0.8 magnitude redder than normal SNe Ia, 
reaching $(B-V)_0 = 1.5$ mag at 10 days after maximum light.
The spectroscopic peculiarity is represented by $i)$ the appearance
of strong Ti~II band in the blue region and $ii)$ the increase of
the ratio of the depths of Si~II absorption troughs at 5800 \AA\ and
6150 \AA\ ($R$(SiII)) (e.g. Nugent et al., 1995). 
Moreover, sub-luminous SNe Ia have steeply declining light curves.
Their decline rate parameter is $\Delta m_{15}(B) > 1.9$ instead of
1.0 $-$ 1.5 as for normal SNe~Ia (e.g. Phillips et al., 1999). 
Because both $R$(SiII) and $\Delta m_{15}(B)$ correlate with the
peak luminosity, the observational data indicate that sub-luminous
SNe Ia have less energetic explosions and less ejected Ni masses 
than normal SNe Ia, although the correlation between the different
quantities cannot be described with a single parameter 
(Hamuy et al., 1996, Hatano et al., 2000).  

In this paper we present medium resolution spectra of the sub-luminous
Ia-type supernova SN~1999by occured in NGC~2841.  Chronicles of the
discovery and the earliest observations can be found in IAU
Circulars No. 7156$-$7159.  Gerardy \& Fesen (1999) pointed out very
soon that the spectrum of SN~1999by is of Type Ia, while Garnavich et al.
(1999) noted that SN~1999by belongs to the group of
sub-luminous SNe~Ia based on the $R$(SiII) line depth ratio. Early-phase
spectra of such SNe are available only for SN~1991bg and SN~1997cn. Thus,
the spectroscopic observations made around maximum light may supply
interesting details about peculiar SNe Ia. Very recently Howell et al.
(2001) presented spectropolarimetric evidence for the strong 
($\approx 20 \% $) asphericity of the ejecta of SN~1999by, suggesting
a connection between the observed asymmetry and the mechanism 
producing sub-luminous SNe Ia.

Our new observations are
described in Sect.2, while their interpretation is given in Sect.3. Sect.4
summarizes the results of this paper.

\section{Spectroscopic observations}

SN~1999by was observed with the 1.88 m telescope and Cassegrain spectrograph 
of the David Dunlap Observatory on three nights between May 10 and 21, 1999. 
The 150A grating in 2nd order with an order separation filter in the
stellar beam was applied, resulting in a reciprocal dispersion of
1.22 \AA/pixel on the CCD detector. The recorded spectra cover about
1250 \AA\ centered on 6000 \AA. 

\begin{figure}
\begin{center}
\psfig{file=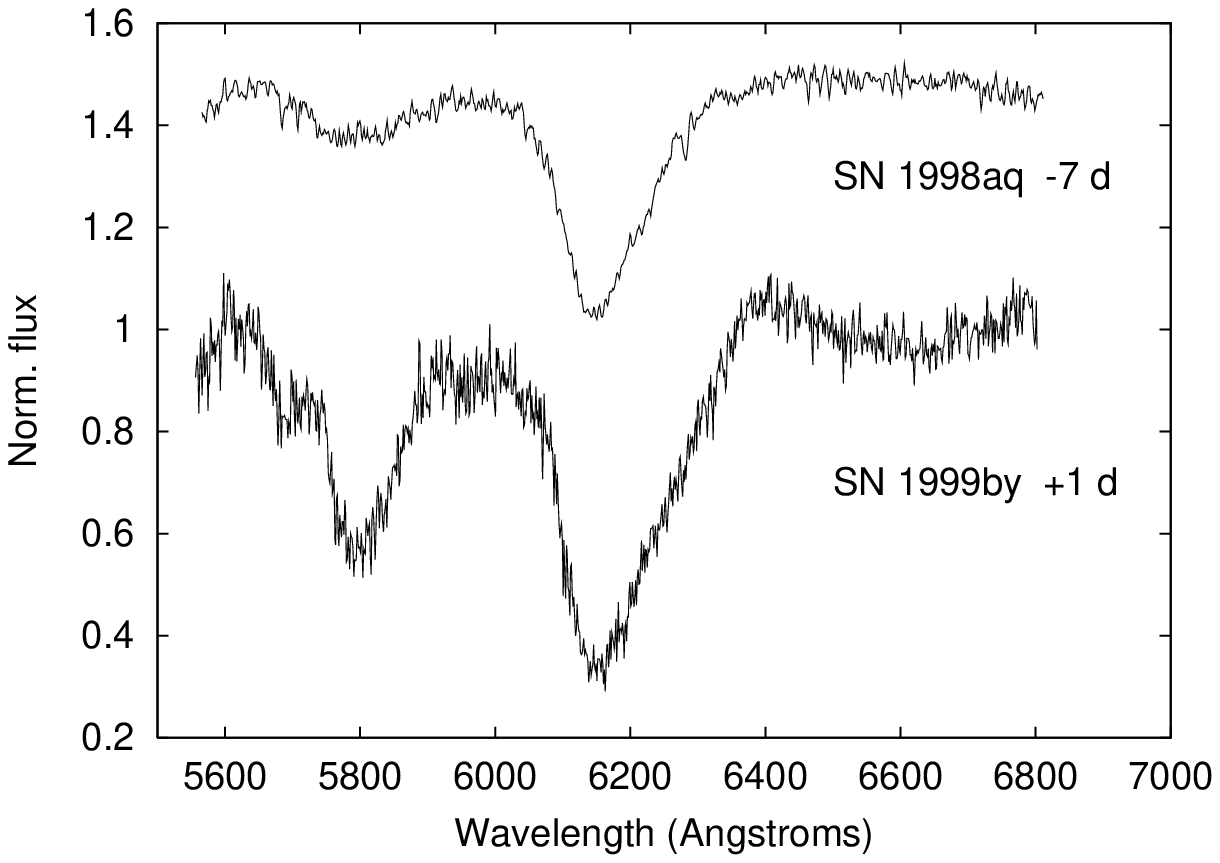,width=8cm,height=7cm}
\caption{Comparison of spectra of SN 1999by (bottom) and SN 1998aq
(top) around maximum light. The latter spectrum was shifted by 0.5. 
The relative strength of the 5800 \AA ~Si~II trough in 
SN 1999by is apparent.}
\end{center}
\end{figure}

Reduction was performed by standard routines in 
$IRAF$\footnote{$IRAF$ is distributed by NOAO which is operated by the
Association of Universities for Research in Astronomy 
(AURA) Inc. under cooperative agreement with the National 
Science Foundation}. The wavelength
scale was calibrated with FeAr spectral lamp exposures.
The background light due to the host galaxy was approximated
with a quadratic polynomial perpendicular to the dispersion axis 
and subtracted from the SN aperture at each wavelength. 

Because of the rarity of photometric nights at the DDO site
(in the proximity of Toronto), precise flux calibration of
the spectra was not possible. However, an approximate 
correction for the CCD spectral response has been determined
by obtaining the spectra of three nearby field stars 
(\#2, \#3, \#4 of Skiff, 1999; see Fig.1 of T\'oth \& Szab\'o, 2000).
Their $(B-V)$ indices (0.59, 0.75, 0.60 mag, respectively) 
corresponds to $T_{eff} \approx 5500 ~-~ 6000$ K according to the
tables of Schmidt-Kaler (e.g. Carroll \& Ostlie, 1996).
The measured continua of these stars could be approximated by a smooth
polynomial. The real continuum fluxes (assuming $\log g = 4$)
were estimated from model spectra of Buser \& Kurucz (1992).
Then, the wavelength-dependent sensitivity of the used CCD
spectrograph was calculated by dividing the measured continua with
the chosen model continuum flux values. The resulting function
turned out to be smooth and slightly decreasing to the red.
We have looked for any sudden drop in the sensitivity function
that might distort the depths of spectral lines and found nothing.

The SN spectra were then corrected by dividing the measured
spectra with the approximate sensitivity function described
above. Because the used comparison stars were not spectroscopic 
standards, the corrected ``continuum'' slopes of the SN spectra
may still be slightly distorted. Nevertheless, the SN spectra 
calibrated in such a way are acceptable
for a comparative analysis described in Sect.3.1.

The reduced spectra are collected and plotted in Fig.1.
The epochs of the observations are indicated on
the right side of each spectrum. The phases relative
to B-maximum were determined by adopting the result of
Bonanos, et al. (1999) that the maximum in $B$ occured
on JD 2451308.5 (May 9/10, 1999), which was also
confirmed by photometry of T\'oth \& Szab\'o (2000)
(note that the JD of maximum light was misprinted
in both papers). 

Although the spectral coverage is rather small, the
appearance of the strong Si~II line around 6150 \AA\ 
unambigously identifies a Type~Ia SN. The other strong 
feature, the absorption trough at 5800 \AA\, is also attributed
to Si~II. The relative strength of this line
to that of the 6150 \AA\ line $R$(SiII) 
(Nugent et al., 1995) suggests that SN~1999by is 
a sub-luminous event, as mentioned in Sect.1. 
This is illustrated in Fig.2, where the maximum-light
spectrum of SN~1999by is plotted together with
a pre-maximum spectrum of SN~1998aq (Vink\'o et al., 1999)
obtained with the same instrument and setup (unfortunately,
no spectrum of SN~1998aq closer to its maximum light was
available for us). The spectra in Fig.2 were normalized
to the local ``continuum'' for illustrative purpose.
It is visible that the 5800 \AA\ line is much weaker
in SN~1998aq than in SN~1999by. This line gets stronger 
in pre-maximum spectra of SNe Ia when approaching the
maximum, however, in ``normal'' SNe the increase of
$R$(SiII) is only 0.04$-$0.06
(Riess et al., 1998b). Thus, the 5800 \AA\ feature was
not much stronger in the spectrum of SN~1998aq at maximum.

\section{Data analysis and interpretation}

\subsection{Line depths and radial velocities}

\begin{table}
\caption{Radial velocities of SN~1999by from the minimum of
the Si~II line. The velocity of NGC 2841 ($v_{gx}=638$ kms$^{-1}$)
has been added to get rest-frame radial velocities of the SN.}
\begin{center}
\begin{tabular}{ccc}
\tableline
\tableline
~JD & $\tau$ & $v_r$ \\
($-$2400000) & (day)  & (kms$^{-1}$) \\
\tableline   
51309.6 & +1 & 10174 \\
51312.5 & +4 & 9560 \\
51319.6 & +11 & 8946 \\
\tableline
\end{tabular}
\end{center}
\end{table}

\begin{figure}
\begin{center}
\psfig{file=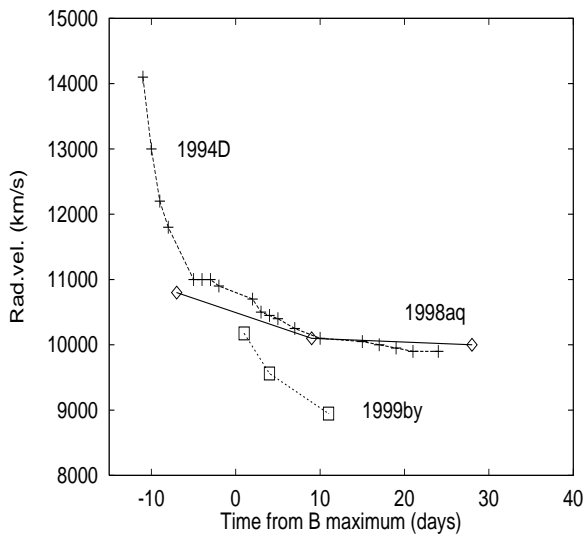,width=8cm,height=7cm}
\caption{Radial velocity curve of SN~1999by computed
from the minimum of the Si~II 6150 \AA\ line 
(boxes), compared with those of SN~1998aq 
(diamonds) and SN~1994D (pluses).
See text for references.}
\end{center}
\end{figure}

We have calculated the value of $R$(SiII) of SN~1999by 
following the definition of Nugent et al. (1995).
$R$(SiII) = 0.63 was obtained for the May 11 spectrum,
while $R$ = 0.66 was determined for the May 14 one,
in accord with the result of Garnavich 
et al. (1999) (see also Hatano et al., 2000).
This is more than a factor of 2 larger than its
usual value of $R = 0.25 \pm 0.1$ for normal SNe Ia
(Riess et al, 1998b, Hatano et al., 2000).

Very recently Howell et al. (2001) published three spectra
of SN~1999by made just before our observations. 
From their Fig.1 we estimated $R$(SiII) = 0.68, 0.70, 0.66 at epochs
$\tau = -2, -1, 0$, respectively. These agree very well
(within a few hundredths) of the results derived from
our spectra, suggesting that our observations probably
do not contain wavelength-dependent systematic errors
distorting the line-depth ratio.

$R$(SiII) correlates with the luminosity parameter
$\Delta$ introduced by Riess et al. (1996) in
the Multi-Color Light Curve Shape (MLCS) method.
A calibration of the linear dependence of $R$(SiII) on $\Delta$ was
given by Riess et al. (1998b), resulting in
$\Delta = 2.33 R - 0.55 (\pm 0.22)$ for $+1$ day phase, while
for $+4$ days the slope and zero point changes to 2.06 and $-$0.46,
respectively. Substituting the $R$(SiII) values given
above one can get $\Delta = 0.90 \pm 0.23$ for the luminosity
parameter of SN~1999by (the given error takes into account 
the quoted uncertainty of the $R$ - $\Delta$ calibration). 
For ``normal'' SNe Ia $\Delta \approx 0$,
therefore, this high value suggests that SN~1999by is one
of the most sub-luminous SNe Ia (Bonanos et al., 1999). 
Note that this spectroscopic $\Delta$ is not suitable 
for describing the multi-color light curve of SN~1999by, 
which requires $\Delta > 1.0$. The light curve decline
parameter, $\Delta m_{15}(B) = 1.87$, is also indicative
of the low luminosity (Bonanos et al., 1999). 

Radial velocities were inferred from the absorption
minima of the 6150 \AA\ Si~II trough
(fitted by a low-order polynomial around minimum).
Table 1 lists the resulted radial velocities, corrected
for the host galaxy center recession velocity
$v_{gx} = 638$ kms$^{-1}$ (as given in the NASA/IPAC 
Extragalactic Database). As the anonymous referee pointed
out, SN~1999by was closer to the edge of NGC~2841 and could 
have significant orbital velocity in addition to the 
center-of-galaxy radial velocity. The rotation velocity
of NGC~2841 at distances greater than 10 kpc from its
center is $\approx 300$ kms$^{-1}$ (Sofue et al., 1999).
Thus, the uncertainty of the
radial velocities is at least $\pm$ 300 kms$^{-1}$
due to the possible orbital velocity as well as the 
large width of the Si~II line. Nevertheless,
this uncertainty is only a few percent of the
Si~II expansion velocity.

Fig.3 compares the radial velocities of SN~1999by with 
those of SN~1998aq (Vink\'o et al.,1999) and SN~1994D
(Patat et al.,1996). The decline of the 
radial velocities of SN~1999by is much stronger than in
the other two cases. SN~1991bg showed similar behavior
(Filippenko et al.,1992), while the other
sub-luminous SNe Ia were not well covered spectroscopically
around maximum. The stronger velocity decline is also in 
good agreement with
the recent result of Hatano et al. (2000) that 
sub-luminous events have lower velocities at 10 days after maximum
than do normal SNe~Ia. 
This may support the hypothesis that
sub-luminous SNe Ia arise from sub-Chandrasekhar mass
progenitors, thus have less energetic explosion than
normal Type Ia events (see Modjaz et al., 2000 for 
recent review of explosion models). 

\subsection{Absolute magnitude and distance}

\begin{figure}[!h]
\begin{center}
\psfig{file=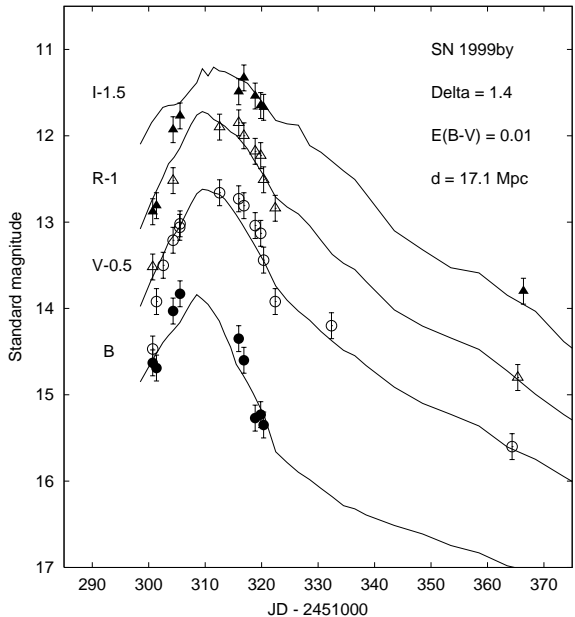,width=8cm}
\caption{MLCS-fit to published $BVRI$ data of SN~1999by.}
\end{center}
\end{figure}

Application of the Multi-Color Light Curve Shape (MLCS) method
for SN~1999by has been published very recently by T\'oth \& Szab\'o
(2000). They used only the $V$ light curve for template fitting
and $(B-V)$ data for the reddening determination, and revealed
$\Delta > 1.6$ mag.
This may indicate that SN~1999by is extremely sub-luminous.
For example, SN~1991bg, the prototype, had $\Delta = 1.44 \pm 0.1$ 
(Riess, Press \& Kirshner, 1996) which is known as one of the
least luminous SNe Ia to date. 

We re-analysed the $BVRI$ photometric dataset published by
T\'oth \& Szab\'o (2000) supplemented by the data of
Hanzl (1999) using the original MLCS template curves of
Riess, Press \& Kirshner (1996). Because the photometric
data are inhomogeneous, systematic errors may distort the
parameters derived from them. Thus, the results below can
be considered only as crude estimates.

Although Riess et al. (1998a)
developed a second version of MLCS (MLCS-2, based on an extended
list of calibrating SNe), the updated template vectors 
have not been published yet. However, recently Vink\'o et al. (2001)
demonstrated that for nearby SNe, even with moderate reddening, 
the MLCS-1 and MLCS-2 methods produce consistent results, if the maximum
magnitude of the fiducial ($\Delta = 0$) light curve is set
properly. We adopted $M_V^{max} = -19.46$ mag for the fiducial
$V$ light curve (Riess, personal communication), while the 
original MLCS-1 method was based on $M_V^{max} = -19.36$ mag. The
difference has been taken into account by adding $+0.1$ mag
to the distance modulus computed with MLCS-1. 

For SN~1999by, $E(B-V) = 0.01$ was adopted,
consistent with the all-sky reddening map of Schlegel et al.
(1998) (T\'oth \& Szab\'o derived $E(B-V) = 0.05 \pm 0.02$ 
from their SN photometry). The template vectors were fitted
simultaneously to all $BVRI$ data, i.e. the residual of all
data were combined in a single $\chi^2$-function. The fitting
was restricted for $\tau > -2$ days, because 
the available pre-maximum data of SN~1999by could not be
adequately fitted together with the post-maximum data
(see also T\'oth \& Szab\'o, 2000). Also, the minimum
photometric error was increased to $\pm 0.15$ mag taking into
account some possible systematic uncertainties in
the used dataset (due to e.g. errors in the standard 
transformation).

\begin{figure}[!h]
\leavevmode
\begin{center}
\psfig{file=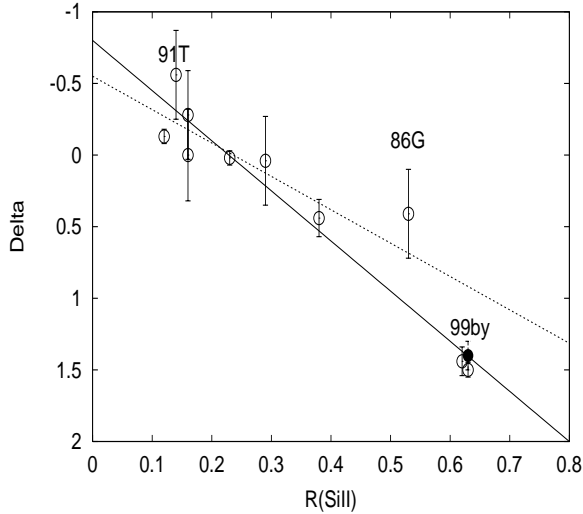,width=8cm,height=7cm}
\end{center}
\caption{Dependence of the light curve parameter $\Delta$ on 
$R$(SiII) using data of well-observed SNe Ia
(see Table 2 for references). The dotted line represents
the relation given by Riess et al. (1998b), while the solid
line shows our new calibration (see text).}  
\end{figure}

The simultaneous fitting of the template curves
resulted in $\Delta = 1.4 \pm 0.1$ and $\mu_0 = 31.16 \pm 0.15$ 
mag for the luminosity parameter and the reddening-free distance
modulus, respectively. The given uncertainty of the distance 
modulus is only the rms error of the fitting, disregarding
the possible systematic errors in the absolute magnitudes of
the template light curves. The quality of the fit can be
judged from Fig.4. It is seen that the fitting is
not perfect, however, we believe that this is close 
the optimal result that can be achieved from the 
present datasets. The minimum of the reduced chi-squared that
could be reached was $\chi^2 \approx 1.5$. Still, the solution
may be systematically incorrect, because for such under-luminous SN
the template vectors in MLCS-1 are based only on the 
light curves of SN~1991bg, and the light variation of SN~1999by
may differ substantially from that. The large deviation of
the pre-maximum points may be due to such inconsistence.
Also, photometric errors higher than we adopted 
($\pm 0.15$ mag) cannot be ruled out.

The photometric $\Delta$ is significantly higher than the 
value found from spectroscopy ($\Delta = 0.90$, Sect.3). 
It was not possible to reach an
adequate fit with $\Delta = 0.9$ given by spectroscopy.
Thus, the current spectroscopic determination
of $\Delta$ does not give a suitable result for SN~1999by,
which is probably true for other extreme sub-luminous SNe~Ia as well.
An improved $R$(SiII) $-$ $\Delta$ calibration is presented 
in the next section. 

Although sub-luminous SNe Ia are usually eliminated from
the sample of ``good'' distance indicators, the MLCS
method gives information on their distances as well.
The new distance modulus of SN~1999by given above corresponds
to $17.1 \pm 1.2$ Mpc geometric distance, which is 
significantly higher than $d = 10-12$ Mpc given by
T\'oth \& Szab\'o (2000) and Tully (1988). Note that
the template vectors of the MLCS-1 method were calibrated
from a training set of SNe adopting Tully-Fisher- and
surface-brightness-fluctuation distances, but the
final absolute magnitudes were shifted
to match the Cepheid-based distance scale (Riess,
Press \& Kirshner, 1996). Thus, the higher distance
given above simply reflects the difference between
various distance scales, and may still contain a
systematic error. Both
Cepheid- and Tully-Fisher distances undergo 
significant revisions nowadays (e.g. Gibson et al., 2000,
Sakai et al., 2000), so this question can be
answered only when the problems of the different
distance scales are solved. 

\subsection{Comparison with other SNe Ia}

The absolute magnitudes of SNe Ia correlate with
$R$(SiII) as was presented very clearly by Nugent et al. (1995).
Although Hatano et al. (2000) demonstrated that this correlation
is not a one-parameter sequence, the dependence of $M_V^{max}$ on
$R$(SiII) may still be a somewhat useful relation for
predicting SNe luminosities from spectra observed around maximum
(at least as a first approximation). Now the correlation can be
slightly improved using the new data of sub-luminous SNe Ia.

\begin{table}[!h]
\caption{$R$(SiII) and $\Delta$ parameters of
SNe~Ia at maximum. }
\begin{center}
\begin{tabular}{lcrcl}
\tableline 
\tableline
SN & $R$(SiII) & $\Delta$ &$\sigma_{\Delta}$ & Ref. \\
\tableline 
1994ae & 0.12 & $-$0.13 & 0.05 & a,b \\
1991T &  0.14 & $-$0.56 & 0.31 & a,b \\
1981B &  0.16 & 0.00 & 0.32 & a,b \\
1990N &  0.16 & $-$0.28 & 0.31 & a,b \\
1998bu & 0.23 & 0.02 & 0.05 & d,e \\
1989B &  0.29 & 0.04 & 0.31 & a,b \\
1992A &  0.38 & 0.44 & 0.13 & a,b \\
1986G &  0.53 & 0.41 & 0.31 & a,b \\
1991bg & 0.62 & 1.44 & 0.10 & a,b \\
1997cn & 0.63 & 1.50 & 0.05 & c \\
1999by & 0.63 & 1.40 & 0.10 & pp \\ 
\tableline
\end{tabular}
\end{center}
References: a) Nugent et al. (1995); b) Riess, Press \& Kirshner (1996);
c) Turatto et al. (1998); d) Jha et al. (2000); e) Hatano et al. (2000);
pp) present paper
\end{table}

We have collected values of the light curve parameter $\Delta$ 
that were derived via the MLCS method and
the values of $R$(SiII), both at $V-$maxi\-mum. These
are listed in Table 2. Because $\Delta$ measures the deviation of
the $V$ light curve of a particular SN from the fiducial SN~Ia
light curve at maximum, it is also directly proportional to
the maximum absolute magnitude $M_V^{max}$. However, at present, 
$\Delta$ seems to be a more useful parameter than 
$M_V^{max}$, because it is determined directly from the
shape of the (multi-color) light curves, therefore it is
not affected by the adopted zero-point of the distance scale.

The result is plotted in Fig.5, where 
the continuous line represents the approximate relation
\begin{eqnarray}
\Delta ~=~ & 3.5 ~R\hbox{(SiII)} & ~-~ 0.8 \\
 ~         & \pm 0.44    ~       & ~\pm 0.17 \nonumber
\end{eqnarray}
while the dashed line shows the relation given by 
Riess et al. (1998b) (see Sect.3.1). It is seen
that the Riess et al. calibration gives systematically
lower $\Delta$ for the most sub-luminous SNe~Ia. 
The new relation fits these SNe better,
thus it may be used to predict $\Delta$ from
the measurement of $R$(SiII) around maximum. Such
prediction may be useful, for example, in the ``snapshot''
distance estimate method (Riess et al., 1998b)
when only a few nights' data are available for
a particular SN.

From Fig.5, it may also be interesting
that the most sub-luminous SN~1991bg, SN~1997cn and
SN~1999by represent a small group with very consistent
line-depth ratio and $\Delta$ (or absolute magnitude). 
It may mean that sub-luminous SNe Ia might be 
better distance indicators
than previously thought, if their sub-luminous status is
properly identified from spectroscopy  and their
difference from normal SNe Ia is taken into account.
It is known (e.g. Hamuy et al., 1994) that such sub-luminous SNe Ia may
be more frequent events than their discovery rate suggests,
thus, it is possible that many more such objects will be detected
in the extensive supernova-search programs. However, the present
sample is obviously too small to draw a certain conclusion.

\section{Summary}

The results of this paper can be summarized as follows:

Three spectra of the sub-luminous Type Ia SN~1999by around maximum
were obtained at DDO, centered on the Si~II 
trough at 6150 \AA. The $R$(SiII) line depth ratio 
(Nugent et al., 1995) is determined to be 0.63 at maximum,
indicative of an intrinsically sub-luminous SNe Ia.

Radial velocities of the 6150 \AA\ Si~II line 
were determined from measuring the Doppler-shift of
the line core. It is shown that SN~1999by exhibited
stronger decline of its radial velocity curve than
other, spectroscopically normal SNe Ia. This may
support the low-energy, sub-Chandrasekhar mass explosion
proposed for such SNe.

Absolute magnitude and distance of SN~1999by have been 
derived via the MLCS method. The assumption that the reddening
is $E(B-V)=0.01$ led to $M_V^{max}$=$-$18.06$\pm 0.1$ mag and
$d = 17.1 \pm 1.2$ Mpc. The maximum absolute 
magnitude as well as 
the line depth ratio $R$(SiII) of SN~1999by are in 
very good agreement with those of other extreme sub-luminous 
SNe Ia, SN~1991bg and SN~1997cn. An approximative
linear relation has been determined to predict the light
curve parameter $\Delta$ of SNe Ia over a wide range of
observed $R$(SiII).

\acknowledgments
This research has been supported by Hungarian OTKA Grant No.T032258,
the Magyary Postdoctoral Fellowship to JV from 
Foundation for Hungarian Education and Science (AMFK), 
the ``Bolyai J\'anos'' Research Scholarship to LLK. 
Thanks are due to the David Dunlap Observatory for allocating
the necessary telescope time. The NASA Astrophysics Data
System, the SIMBAD database and the Canadian Astronomy Data
Centre were used to access data and references. 
The availability of these services are gratefully acknowledged.

\end{document}